\documentclass[
    prl, twocolumn, superscriptaddress, nofootinbib, amsmath, amssymb,
    aps, floatfix
]{revtex4-2}

\usepackage{graphicx,subfigure}
\usepackage{svg}
\svgsetup{
    inkscapepath=i/svg-inkscape/
}
\svgpath{{svg/}}
\usepackage{xcolor}
\usepackage{setspace}
\usepackage{hyperref}

\usepackage[utf8]{inputenc}
\usepackage[T1]{fontenc}
\usepackage{lmodern}

\usepackage{booktabs}
\usepackage{color}
\usepackage{epsfig}
\usepackage{ifpdf}
\usepackage{amsmath}
\usepackage{bm}
\usepackage[english]{babel}
\usepackage{amsfonts}
\usepackage{amssymb}
\usepackage{braket}
\usepackage{enumerate}
\usepackage{cancel}
\usepackage{multirow}
\usepackage{cleveref}
\usepackage{xspace}
\usepackage{array}
\usepackage{fancyvrb}
\usepackage{fontawesome}
\usepackage{dcolumn}
\usepackage{slashed}
\usepackage[normalem]{ulem}
\usepackage{url}

\def\XXint#1#2#3{{\setbox0=\hbox{$#1{#2#3}{\int}$}
     \vcenter{\hbox{$#2#3$}}\kern-.5\wd0}}

\makeatletter
\g@addto@macro\bfseries{\boldmath}
\makeatother

\definecolor{nicered}{rgb}{0.7,0.1,0.1}
\definecolor{nicegreen}{rgb}{0.1,0.5,0.1}
\hypersetup{colorlinks, urlcolor=blue, citecolor=nicegreen,linkcolor= nicered}

\begin{document}
\title{Dark Photon Oscillations in Waveguide} 

\author{Yu-Xin Tian}
\email{tianyx2024@emails.bjut.edu.cn}
\affiliation{Beijing University of Technology, 100124, Beijing, China}

\author{Wenyu Wang}
\email{wywang@bjut.edu.cn}
\affiliation{Beijing University of Technology, 100124, Beijing, China}

\author{Wen-Na Yang}
\email{yangwenna22@mails.ucas.ac.cn}
\affiliation{Department of Physics and Institute of Theoretical Physics, Nanjing Normal University, Nanjing, 210023, China}

\author{Bin Zhu}
\email{zhubin@mail.nankai.edu.cn}
\affiliation{School of Physics, Yantai University, Yantai, 264005, China}

\begin{abstract}
Dark photons, which can kinetically mix with ordinary photons, represent 
the simplest extension to the standard model. Detecting their
oscillations with visible photons could provide crucial insights into 
the nature of dark matter and fundamental interactions beyond the standard 
model. We propose a novel laboratory-based approach to detect dark photon 
oscillations using a laser in an
Optical Time-domain Relectometry (OTDR) setup. The laser light 
propagating through the optical fiber undergoes oscillations with the dark 
photon, leading to measurable changes in the power flow. These oscillations can precisely measured, leveraging its high sensitivity and efficiency in detecting 
small variations in the optical signal. This approach could provide a new avenue for probing dark photon
oscillations in the laboratory and greatly improve the current
experimental sensitivity to dark photon in a wide mass range.


\end{abstract}

\maketitle
\newpage

\section{Introduction}
Dark matter (DM) is a cornerstone of modern physics, confirmed by
cosmological and astrophysical observations~\cite{Rubin, Planck, Anderson:2013zyy, LSS}. Among DM candidates, 
ultralight dark photons~\cite{Holdom:1985ag, Dienes:1996zr, Abel:2003ue, Abel:2006qt}, denoted  $A^{\prime}$, stand out as particularly
compelling.  A dark photon (DP) is a hypothetical vector boson 
associated with a
$ \text{U}(1) $ gauge symmetry in a hidden sector. It can serve as a dark 
matter candidate if it is stable and abundant from the early universe 
processes, such as through inflationary fluctuations 
\cite{Graham:2015rva, Ema:2019yrd, Kolb:2020fwh, Salehian:2020asa, Ahmed:2020fhc, Nakai:2020cfw, Nakayama:2020ikz, Kolb:2020fwh, Salehian:2020asa, Firouzjahi:2020whk, Bastero-Gil:2021wsf, Firouzjahi:2021lov, Sato:2022jya}, parametric resonances \cite{Co:2018lka, Dror:2018pdh, Bastero-Gil:2018uel, Agrawal:2018vin, Co:2021rhi, Nakayama:2021avl, Cyncynates:2023zwj}, the decay of cosmic strings \cite{Long:2019lwl} or
misalignment mechanisms~\cite{Nelson:2011sf, Arias:2012az, AlonsoAlvarez:2019cgw, Nakayama:2019rhg, Nakayama:2020rka}. 
Ultralight masses, spanning $10^{-22}$ to
$10^{-1}$ eV, enable wave-like behavior, addressing astrophysics' 
small-scale structure issues~\cite{Hui:2016ltb,Suarez:2013iw, Review_SFDM_Arturo}. 
Moreover, dark photons can kinetically mix with ordinary photons, $\gamma$, 
through a renormalizable interaction $\varepsilon 
F^{\mu\nu}F_{\mu\nu}^{\prime}$~\cite{Holdom:1985ag} that naturally 
arises from the blending of 
the two $ \text{U}(1) $ gauge fields. This mixing leads to oscillations
between photons and dark photons, similar to neutrino oscillations,
providing a promising new avenue for detection. 

Dark photons are typically probed through their conversion to visible 
photons, producing a resonant signal in detectors. This assumes DPs are 
present in the laboratory, either as ambient dark matter or as emissions 
from the Sun.  Existing methods for DPs include haloscope 
experiments like TOKYO~\cite{Suzuki:2015vka, Suzuki:2015sza, Knirck:2018ojz, Tomita:2020usq}, FUNK \cite{FUNKExperiment:2020ofv}, 
DM pathfinder and Dark E-field \cite{Phipps:2019cqy, Godfrey:2021tvs}, SHUKET \cite{Brun:2019kak}, WISPDMX \cite{Nguyen:2019xuh}, SQuAD \cite{Dixit:2020ymh}, 
light shining through 
walls~\cite{DePanfilis:1987dk, Hagmann:1990tj, Gelmini:2020kcu, Caputo:2021eaa}, plasma telescopes~\cite{Lawson:2019brd, Gelmini:2020kcu}, Dark E-field \cite{Godfrey:2021tvs}, DM-Radio~\cite{Parker:2013fxa, Chaudhuri:2014dla, 7750582, Chaudhuri:2018rqn}, MADMAX~\cite{Caldwell:2016dcw}, CMB spectrum distortions~\cite{Arias:2012az,  McDermott:2019lch, Caputo:2020bdy, Witte:2020rvb},
and radio observations of solar emissions~\cite{An:2020jmf, An:2023wij}. 

In contrast, our method does not rely on the assumption that 
DPs constitute the dark‑matter component: even if they serve 
merely as virtual particles predicted by new physics, 
they remain directly testable within our 
framework. In this Letter, we propose a laboratory-scale approach to detect
DPs using a waveguide (optical fiber) setup.
The core mechanism involves the 
oscillation between DPs and visible photons, 
driven by an effective mass term in specific environments. 

This effective mass naturally emerges in
waveguides as distinct propagation modes, or in cavities as resonant modes.
Unlike a cavity, where a single photon is measured and 
enhanced by the dissipation rate or quality factor $Q$, oscillations in 
waveguides occur during photon propagation. In the absence of 
DP-photon mixing, photon propagation is analogous to an overdamped
harmonic oscillator: its intensity decays exponentially with distance 
$P\sim  P_0\exp(-\beta z)$~\cite{Kapron:1970,Miya:1979}.
When mixing is present, however, the system 
behaves like an underdamped oscillator, exhibiting spatial oscillations in
the photon power. In this regime, the signal can drop to zero and
subsequently revive to a smaller amplitude - a smoking gun of oscillation 
$P\sim P_0\exp(-\beta z)\cos(\Delta k z)$. By searching for this 
underdamped behavior, we gain direct sensitivity to DP.

Varying the radius of the optical fiber can tune 
the propagation modes to have resonant mixing with the DP, giving us
a scan method. Then the smoking gan can be detected by measuring the
power flow in Optical Time-domain reflectometry 
(OTDR).\cite{Kapron:72,Barnoski:76} 
By comparing the observed spatial variation of the fitted Lorentzian
amplitude to its statistical uncertainty in each resolution bin, 
this approach can achieve up to five orders of magnitude improvement 
in \(\varepsilon\) over existing methods, thereby offering a powerful
and complementary means to explore dark-photon parameter 
space alongside cosmological searches.

\section{Dark Photon Propagation and Oscillation}
Dark photons, hypothetical particles from a dark $U(1)$ gauge theory, 
couple to visible photons via kinetic mixing, described by the Lagrangian
\begin{equation}
\mathcal{L} = -\frac{1}{4} F'_{\mu\nu} F'^{\mu\nu} - \frac{1}{2} m_{A'}^2 
A'_\mu A'^\mu - \frac{\varepsilon}{2} F'_{\mu\nu} F^{\mu\nu},
\end{equation}
where $\varepsilon \ll 1$ is the mixing parameter, $m_{A^{\prime}}$ is the 
dark photon mass, $F_{\mu\nu}^{\prime}=\partial_{\mu}A_{\nu}^{\prime}-
\partial_{\nu}A_{\mu}^{\prime}$ is the dark photon field strength and 
$F_{\mu\nu}$ is the visible photon field strength. This kinetic mixing 
enables oscillations between visible and dark photons 
when their effective masses align. 

In waveguides, such as hollow metallic cylinders or optical fibers,
electromagnetic waves are transversely confined, supporting only 
discrete transverse modes (TE or TM) with cutoff frequencies $\omega_\xi$. 
Frequencies below this cutoff do not propagate, behaving like massive 
photons with effective mass $m_\xi=\omega_\xi$, yielding 
$k_z^2 = \omega^2 - m_\xi^2$. Unlike cavities, where standing waves 
lack axial propagation, waveguides support traveling waves, enabling 
oscillations over distance, a feature absent in cavity-based systems.
The coupled wave equations are
\begin{equation}
\left[ -\frac{\partial^2}{\partial t^2} + \frac{\partial^2}{\partial z^2} 
- \begin{pmatrix} m_\xi^2 & -\varepsilon m_{A^{\prime}}^2 \\ 
-\varepsilon m_{A^{\prime}}^2 & m_{A^{\prime}}^2 \end{pmatrix} \right]
\begin{pmatrix} A \\ A^{\prime} \end{pmatrix} = 0.
\end{equation}
The inclusion of the $\partial^2/\partial z^2$ term enables oscillatory
behavior along the $z$–axis.
The off-diagonal terms, proportional to $\varepsilon m_{A'}^2$, mix the
fields. Diagonalizing the mass matrix via a rotation $B_1 = \cos\theta A -
\sin\theta A'$, $B_2 = \sin\theta A + \cos\theta A'$, with $\tan 2\theta =
2\varepsilon m_{A'}^2 / (m_{A'}^2 - m_\xi^2)$, yields resonant mixing when
$m_\xi \approx m_{A'}$, driving power oscillations along the waveguide.

At the input end ($z=0$) of waveguide, a pure visible photon state 
($A'(0) = 0$) sets $B_1(0) = \cos\theta A(0)$ and 
$B_2(0) = \sin\theta A(0)$. Each eigenstate evolves as 
$B_i(z) = B_i(0) e^{i(k_{z,i}z - \omega t)}$, and the visible field at the
distance $z$ is $A(z) = \cos\theta B_1(z) + \sin\theta B_2(z)$. The power
$P(z)$ includes an interference term 
$\propto \cos[(k_{z,1} - k_{z,2}) z]$, given by
\begin{equation}
P(z)=P_0 e^{-2 \beta z}\left[\eta_1^2+\eta_2^2+2 \eta_1 \eta_2 \cos \left(\Delta k_z z\right)\right]\,.
\end{equation}
with $\eta_1,~\eta_2$ are the ratios of the corresponding
eigenstates.
At resonance ($m_\xi = m_{A'}$), maximal mixing occurs 
($\eta_1 = \eta_2 = 0.5$), yielding 
$P(z) \propto [1 + \cos(\Delta k_z z)]/2$ (up to damping). 


\begin{figure}[htbp]
\centering
\scalebox{0.46}{\includegraphics{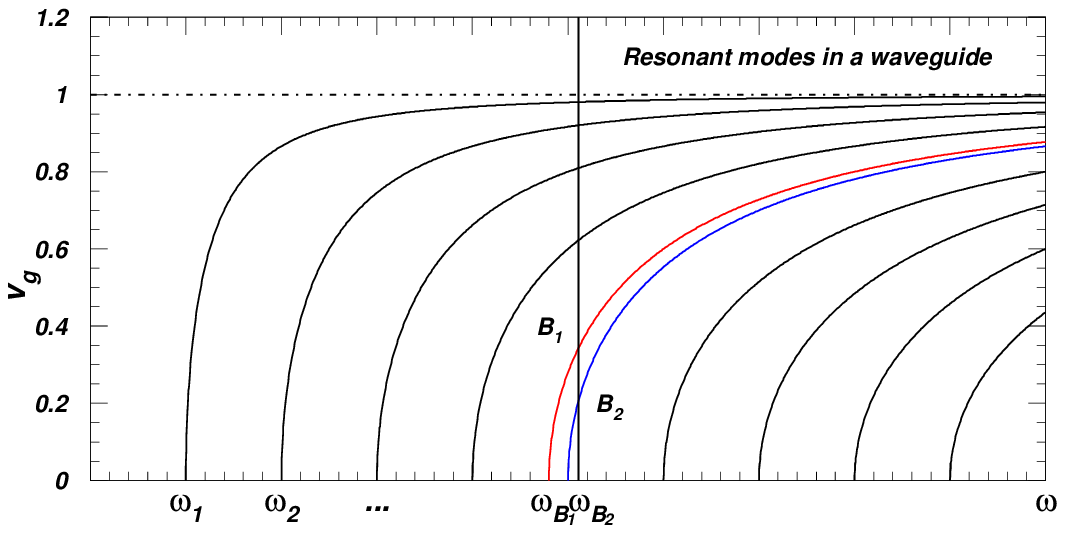}}\\
\scalebox{0.46}{\includegraphics{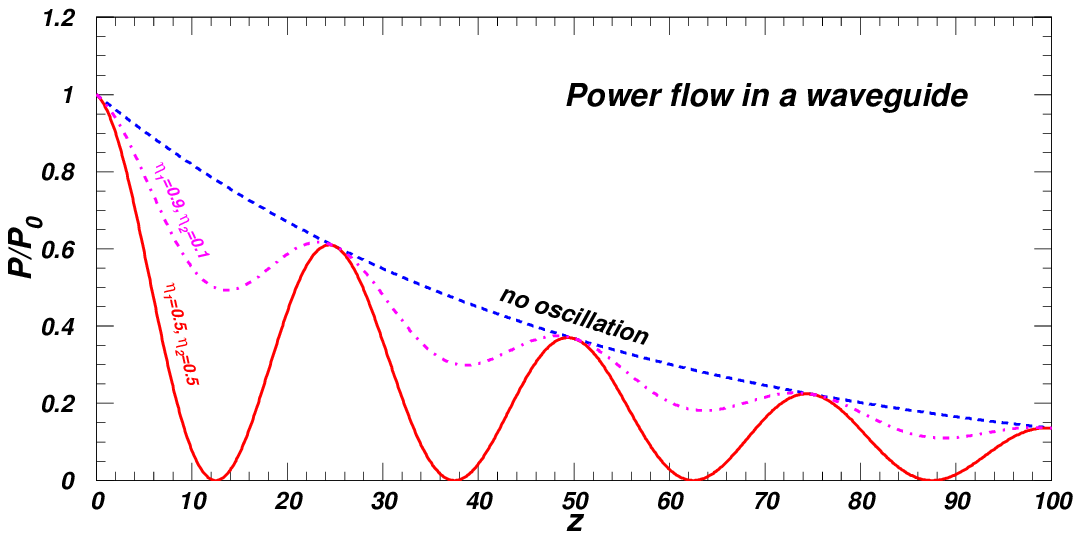}}
\caption{ Dark photon oscillation. Top panel:
sketch of the propagation modes 
in a waveguide. $v_g$ is the group velocity. 
The red and blue curves show the resonant modes by 
the mixing between one mode and dark photon. 
Two different cut off frequency $\omega_{B_1}$, 
$\omega_{B_2}$ are got after the diagonalizaiton.
The vertical line is the input frequency for the scan 
in the proposed measurement.
 Bottom panel: Power flow (normalized by $P_0$) 
along the ($z$) axle. Blue dash curve shows the 
attenuation without oscillation, Magenta dot dash line shows the 
attenuation with oscillation when $\eta_1=0.9,\eta_2=0.1$.
Red solid curve shows the attenuation with oscillation at the resonance,
$\eta_1=0.5,\eta_2=0.5$.}\label{fig1}
\end{figure}

The top panel of Fig.~\ref{fig1}
shows the group velocity of the propagation modes of a waveguide.
In case of the input frequency near the resonance, the phase difference
$\Delta k_z z$ caused by the two orthogonal modes 
will become significant. This will be a different phenomenon from the 
multimode propagation in the waveguide or fiber in which the power
flow will be constant for a monochromatic wave input. 
The oscillation happens in case of the mixing with the dark photon, 
as for that only the visible photon can be detected in the lab.
As shown in bottom panel of Fig.~\ref{fig1},
the blue curve depicts the conventional photon
intensity in the waveguide, exhibiting a purely exponential decay
characteristic of an overdamped oscillator.  Once photon–dark‑photon mixing 
is introduced, energy is periodically exchanged between the two states,
producing the underdamped oscillations illustrated by the red and purple 
curves.  In this regime, the photon power falls to zero at the nodes of the 
oscillation and is subsequently restored to its peak value at the 
antinodes. The oscillation amplitude grows with the mixing parameter, 
becoming significant at maximal mixing, $\eta = 0.5$, where the revival 
of the photon signal is sufficiently large to enable detection of the 
dark photon. 

In contrast to detection schemes that rely on ambient or solar‐sourced dark 
photons, our waveguide–based setup induces photon–dark‑photon mixing in a 
controlled laboratory environment—akin to reactor neutrino production—so 
that no preexisting dark‑photon population is required.  Consequently,
there is no need for the dark photon to occupy a coherent state in the 
initial configuration, a crucial relaxation of the usual assumption.  
Nearly all existing searches presuppose a coherent‐state dark photon 
background to render the conversion signal detectable, yet the coherence of 
any dark‑photon field remains unverified and may therefore challenge the 
viability of such approaches.  By circumventing the coherent‐state 
requirement, our method offers a model‑independent pathway to probe dark 
photons on a tabletop experiment.  

\section{Expected Sensitivity}\label{sec3}

We propose to detect oscillations of visible photons into dark-photon 
states during propagation in an optical fiber, The fibers
are ideal distributed
sensors: they span kilometers, can be embedded readily within structures, 
are immune to electromagnetic interference, and allow each short segment to
act as an independent sensing node.~\cite{culshaw1988fiber} 
When the effective mass of a guided 
mode matches the DP mass, resonant mixing splits the propagation
eigenstates with distinct phase velocities. Interference between these
eigenstates therefore produces a spatially oscillating power flow-rather
than the usual monotonic attenuation-along the guide. OTDR, which measures
position-resolved backscatter over long distances, is a natural and
sensitive technique to detect the resulting cosine-like modulation of
normalized power versus propagation length.

\begin{widetext}
\begin{center}
\begin{figure}[htbp]
\scalebox{1}{\epsfig{file=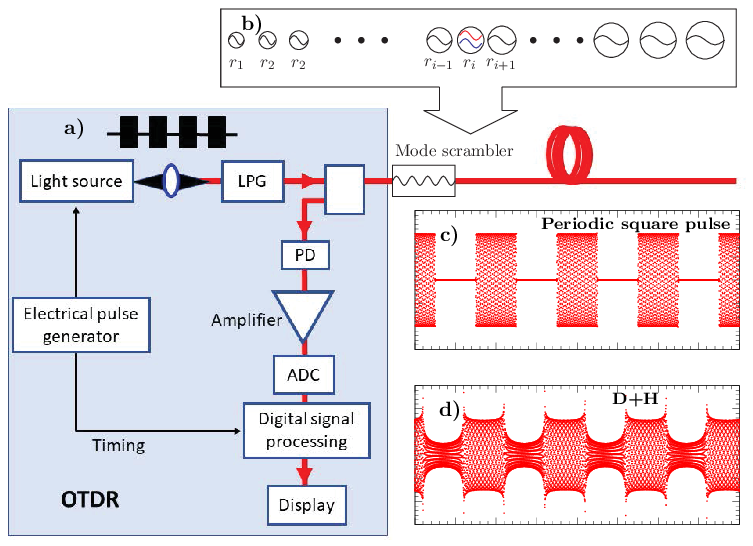}}
\caption{The sketch map of the experimental setup
for measuring the oscillation of DPs by OTDR.}\label{fig2}
\end{figure}
\end{center}
\end{widetext}

Fig.~\ref{fig2} presents the proposed OTDR-based measurement:
\cite{O'Sullivan2023fiber}
\begin{itemize}
\item Panel a): the experimental layout. 
\end{itemize}
A periodic optical pulse from a laser diode, driven by an electrical pulse
generator, is launched into the test waveguide. A long-period grating (LPG) 
selects the guided mode of interest; an optional mode scrambler stabilizes 
the modal power distribution and suppresses cladding modes. Backscattered 
light is routed via a directional coupler to a photodiode (PD), amplified, 
and digitized by an analog–to–digital converter (ADC); digitized traces 
are processed in a DSP that is synchronized to the source pulses so that
the propagation delay of each backscattered feature is determined
precisely. 
\begin{itemize}
    \item Panel b): resonant-mode scan.
\end{itemize}
The simplest way to scan the propagation modes is to prepare 
optical fibers with increasing radii and measure
the corresponding power flow. The oscillation behavior appears
when the effective guided-mode (cut off frequency) mass lies near 
the dark-photon mass $m_{A^{\prime}}$.  Also the 
input frequency of laser should be tuned near the cut off frequency, too.
\begin{itemize}
    \item Panel c): input squared pulse; Panel d): observed signature.
\end{itemize}
The illustrative square pulse ( shown in 
panel c)) is a carrier at $\omega_0$
modulated by a slow envelope $\Omega\ll \omega_0$
and decomposes into the dominant tone $\omega_0$, lower harmonics 
$\{\omega_0-n\Omega\}$, and higher harmonics $\{\omega_0+n\Omega\}$. 
This decomposition motivates the ``D+H'' strategy: 
when $\omega_0$ is tuned near the resonant mode the lower harmonics
become evanescent and are filtered out by the guide, leaving the 
dominant tone plus higher harmonics (shown in panel d))
as the effective launched field. 
Note that there are some distortions of the waveform after a long
propagation in the fiber, which will be shown in the 
supplemental material.
Each spectral component propagates with phase velocity 
$v_{p, B_i}^n=\left(\omega_0+n \Omega\right) / k_{z, B_i}$. 
So the two eigenstates accumulate a component–dependent phase difference
\begin{equation}
    \Delta \varphi_n = 
    L\left(\sqrt{(\omega_0+n\Omega) ^2-m^2 _{B_1}}
    -\sqrt{(\omega_0+n\Omega)^2-m^2_{B_2}}\right),
   \label{phshift}
\end{equation}


Finite laser coherence (fractional linewidth $\zeta=\Delta\nu/\nu$) 
broadens the source and reduces the averaged phase difference. Modeling
each spectral component with a Lorentzian lineshape $g(\omega,\omega_c)$
leads to an averaged phase
\begin{equation}
\overline{\Delta\varphi_0}
=\frac{4\varepsilon L}{\lambda}
\int_{0}^{\infty}\frac{1}{\sqrt{x\zeta+2\varepsilon}+\sqrt{x\zeta}}\frac{1}{1+x^{2}}dx,
\label{phshift}
\end{equation}
Physically, finite $\zeta$ partially washes out the phase accumulated by
the carrier and thus imposes a lower bound on the observable
$\overline{\Delta\varphi_0}$. 

\begin{figure}[htbp]
\begin{center}
\includegraphics[width=0.48\textwidth]{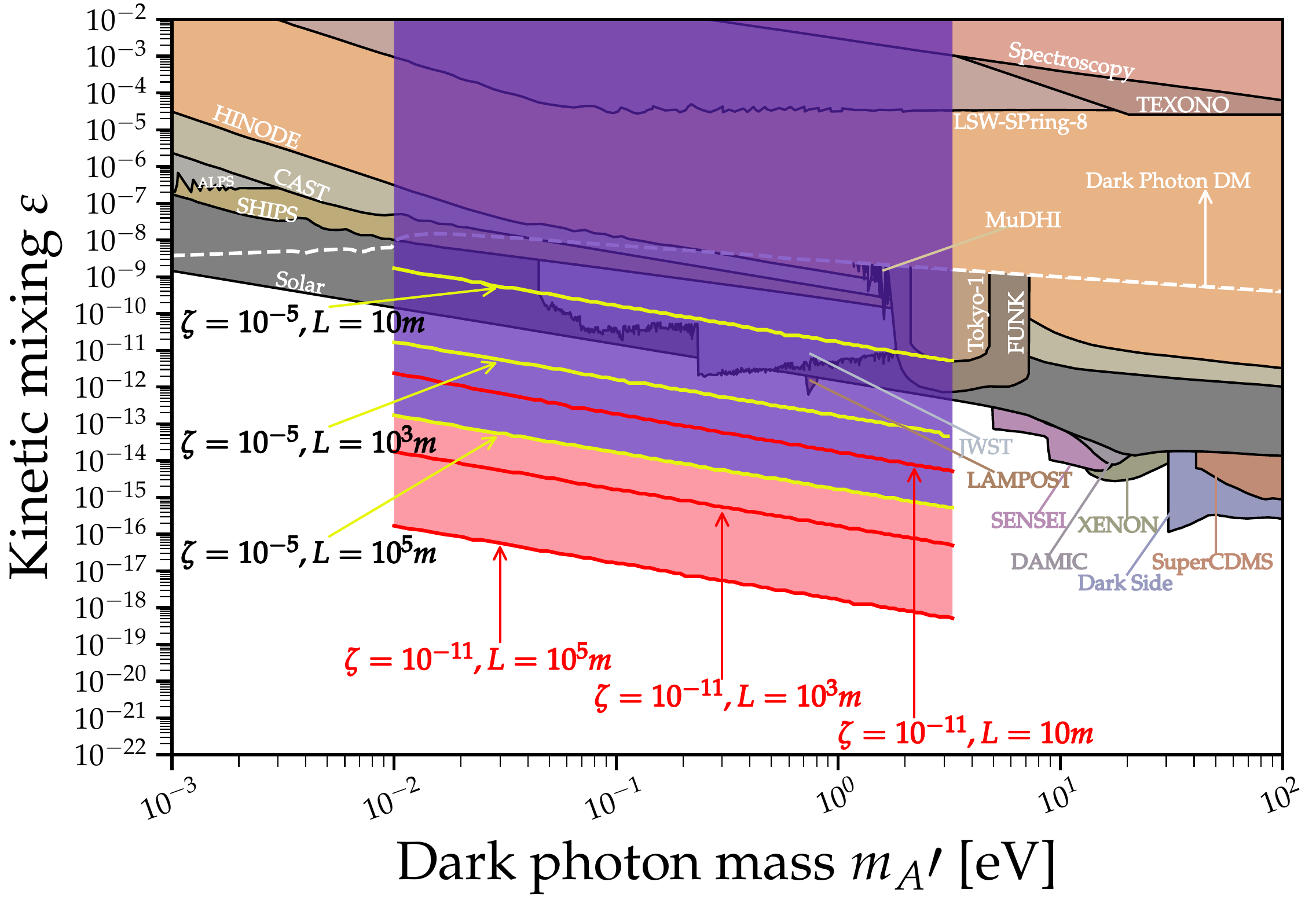}
\caption{Expected sensitivity of the OTDR  to dark photons.
The frequency of the scan is chosen around that of infrared light 
$0.8 {\mu\rm m}-1.7 {\mu\rm m}$ used for 
telecommunication.~\cite{Senior2009fiber}
$\zeta$ is chosen as $10^{-5},~10^{-11}$  for
ordinary laser (yellow lines)  and well controlled ultra-stable
laser (red lines), respectively.~\cite{Thyagarajan2011laser}
And the lengths of the optical 
fiber are set as $10\rm m$,  $10^{3}\rm m$, and $10^{5}\rm m$.
Other limits are got from Ref.~\cite{Caputo:2021eaa}}
\label{fig3}
\end{center}
\end{figure}

In all, the sensitivity in the OTDR
is set by the fit uncertainty per measurement point,
the coherence (linewidth) of the laser, 
and the effective baseline \(L_{\rm eff}\): 
long, coherent fibers and narrow laser linewidths increase accumulated
phase difference and hence reach, whereas finite coherence and practical
noise set the achievable limit on the kinetic-mixing parameter.  This
approach therefore converts distributed, laboratory-scale optical sensing
into a quantitative bound on dark-photon mixing without accessing the 
fiber interior or using auxiliary fields, while requiring only standard
fiber-optic instrumentation and mode control.
In the experiment, we can measure backscatter-derived power $\hat P(z_i)$ 
at a set of lengths  ${\cal M}={\hat P(z_1),\ldots,\hat P(z_n)}$. 
Then we can compute
normalized values $P(z_i)=\hat P(z_i)e^{2\beta_\xi z_i}$ where $\beta_\xi$
is the measured attenuation coefficient, which removes monotonic loss so
that residual spatial structure reveals oscillation.
Fit $P(z)$ to a cosine model (or sinusoidal envelope appropriate to the 
two-eigenstate interference). A significant cosine component with the
expected spatial period is evidence for oscillation. When 
$\lambda_{\rm osc}\gg L$ expand the cosine and use three-point 
measurements $P(0),P(L/2),P(L)$ to estimate $\Delta k_z L$ via
newly defined quantity ${\cal E}(L)$
\begin{equation}
{\cal E}(L)=\frac{P(L/2)-P(0)}{P(L)-P(0)}\simeq \frac{1}{4}\Big[1+\frac{1}{16}(\Delta k_{z}L)^{2}\Big],
\label{eq:extkz_repeat}
\end{equation}
which yields $\Delta k_z L\equiv\overline{\Delta\varphi_0}$. 
Generally, ${\cal E}(L)$ can be measured much more precisely with current fiber technology. For example, the limits in Fig.~\ref{fig3} are set by $({\cal E}-0.25)\times100 \simeq 0.006$. Smaller values of ${\cal E}(L)$ lead to stronger constraints on $\varepsilon$.


As shown in Fig.~\ref{fig3}, numerical results indicate that for common diode or telecom lasers with $\zeta = 10^{-5}$, even a $10~\rm m$ optical fiber can set constraints comparable to the current solar global limit. Extending the baseline to $1\ \rm{km}$ improves the sensitivity by approximately an order of magnitude. Notably, for frequency-stabilized sources with $\zeta = 10^{-11}$, we find that the exclusion limits are enhanced by up to $\mathcal{O}(7)$ orders of magnitude 
across the considered infrared wavelength range ($0.8\text{–}1.7\ \mu\rm{m}$), significantly surpassing existing experimental bounds. 
These results demonstrate that our method opens a new avenue for probing dark photons in previously inaccessible parameter regimes. Fully exploiting this potential requires a trade-off between increasing the baseline $L$ and improving the source coherence.

\section{Conclusions}
We have proposed a resonant, waveguide-based search for oscillations
between visible photons and dark photons and shown that standard optical
components—long-period gratings for mode selection, mode scramblers,
synchronized OTDR, and DSP—readily convert a tiny kinetic mixing into a
macroscopic, position-dependent signature. It can improve sensitivity by
many orders of magnitude over the current probe. 
The proposal is conceptually simple, experimentally tractable, and 
complementary to existing cavity, haloscope, and light-shining-through-wall
searches: it probes a kinematic oscillation phenomenon that redistributes
power between visible and invisible sectors rather than relying on ambient
dark-photon flux. Practical challenges—mode-resolution in fabricated
fibers, laser coherence, and enhanced attenuation near cutoff—are
identified and mitigations are provided.

\section{Acknowledgement}
We would like to thank Lei Wu, Xiao-Hui Fang, Yun-Xin Wang, Qi-Wei Liang, Ji-Heng Guo and Yu-Song Cao for useful discussions. This work is supported by
the NSFC under Grants No. 12475104, No. 12275232, No. 12335005, and No. 12275134.

\bibliography{refs}

\onecolumngrid
\clearpage

\setcounter{page}{1}
\setcounter{equation}{0}
\setcounter{figure}{0}
\setcounter{table}{0}
\setcounter{section}{0}
\setcounter{subsection}{0}
\renewcommand{\theequation}{S.\arabic{equation}}
\renewcommand{\thefigure}{S\arabic{figure}}
\renewcommand{\thetable}{S\arabic{table}}
\renewcommand{\thesection}{\Roman{section}}
\renewcommand{\thesubsection}{\Alph{subsection}}

\newcommand{\ssection}[1]{
    \addtocounter{section}{1}
    \section{\thesection.~~~#1}
    \addtocounter{section}{-1}
    \refstepcounter{section}
}
\newcommand{\ssubsection}[1]{
    \addtocounter{subsection}{1}
    \subsection{\thesubsection.~~~#1}
    \addtocounter{subsection}{-1}
    \refstepcounter{subsection}
}
\newcommand{\fakeaffil}[2]{$^{#1}$\textit{#2}\\}

\thispagestyle{empty}
\begin{center}
    \begin{spacing}{1.2}
        \textbf{\large
            \hypertarget{sm}{Supplemental material:} Dark Photon Oscillations in Waveguide}\\
    \end{spacing}
    \par\smallskip
    Yu-Xin Tian,$^{1}$
    Wenyu Wang,$^{1}$
    Wen-Na Yang,$^{2}$
    Bin Zhu$^{3}$
    \par
    {\small
       \fakeaffil{1}{Beijing University of Technology, Beijing, China}
        \fakeaffil{2}{Department of Physics and Institute of Theoretical Physics, Nanjing Normal University, Nanjing, 210023, China}
        \fakeaffil{3}{School of Physics, Yantai University, Yantai 264005, China}
    }
\end{center}
\par\smallskip

This section provides a step-by-step derivation that establishes the theoretical foundation for dark photon oscillation, detailing its mechanism, necessity, and how it guides the experimental probe.

\section{Maxwell’s Equations and  Propagation Modes in a Waveguide}

We start from Maxwell’s equations in a uniform, lossless medium of permittivity $\epsilon$ and permeability $\mu$, inside a hollow conducting waveguide. We assume time-harmonic fields $\propto e^{-i\omega t}$. Maxwell’s curl equations give,
\[
\nabla \times \mathbf{E} = i\omega \mathbf{B}, 
\quad
\nabla \times \mathbf{B} = -i\mu\epsilon\omega \mathbf{E},
\quad
\nabla\cdot\mathbf{E}=0,\quad \nabla\cdot\mathbf{B}=0.
\]
We seek solutions propagating along the $z$-axis: 
\[
\mathbf{E}(x,y,z,t) = \mathbf{E}_t(x,y)\,e^{i k_z z - i\omega t}, 
\quad
\mathbf{B}(x,y,z,t) = \mathbf{B}_t(x,y)\,e^{i k_z z - i\omega t}.
\]
Substituting into Maxwell’s equations yields the transverse Helmholtz equation:
\begin{equation}\label{eq:transverse-Helmholtz}
\bigl[\nabla_t^2 + (\mu\epsilon\,\omega^2 - k_z^2)\bigr]\begin{pmatrix}\mathbf{E}_t \\ \mathbf{B}_t\end{pmatrix} = 0,
\quad \nabla_t^2 \equiv \frac{\partial^2}{\partial x^2} + \frac{\partial^2}{\partial y^2}.
\end{equation}
By boundary conditions on the metallic walls (perfect conductor 
idealization first), only discrete transverse modes $(\mathbf{E}_t,\mathbf{B}_t)$ satisfying appropriate vanishing
tangential fields exist.

Eq.~\eqref{eq:transverse-Helmholtz} identifies allowed transverse mode 
functions and defines a cutoff condition. Each mode behaves as if the
photon has an effective “mass” $m_\xi$ given by its cutoff frequency. 
For each mode label $\xi$, the transverse eigenvalue equation yields
\begin{equation}
\nabla_t^2 \psi_\xi + \mu\epsilon\,\omega_\xi^2 \,\psi_\xi = 0,
\end{equation}
with cutoff frequency $\omega_\xi$ determined by geometry (e.g., for rectangular waveguide TE$_{mn}$ mode, $\omega_{mn}^2 = \pi^2(\tfrac{m^2}{a^2} + \tfrac{n^2}{b^2})/(\mu\epsilon)$). Define effective mass
$m_\xi \equiv \omega_\xi$, then propagation along $z$ obeys
\begin{equation}
k_z^2 = \omega^2 - m_\xi^2.
\end{equation}
Hence, the wave equation for the field envelope along $z$ is analogous to a Klein–Gordon equation,
\begin{equation}
\bigl[-\partial_t^2 + \partial_z^2 - m_\xi^2\bigr]\psi = 0.
\end{equation}
Phase and group velocities follow
\begin{equation}\label{eq:vp-vg}
v_{p,\xi} \;=\; \frac{\omega}{k_z} = \frac{\omega}{\sqrt{\omega^2 - m_\xi^2}}, 
\quad
v_{g,\xi} \;=\; \frac{d\omega}{dk_z} = \sqrt{1 - \frac{m_\xi^2}{\omega^2}}, 
\quad v_{p,\xi} v_{g,\xi} = 1.
\end{equation}

Viewing each mode as a “massive photon” clarifies how a dark photon of mass $m_{A'}$ can resonantly mix when $m_{A'} \approx m_\xi$. We briefly recall the separation into TE and TM modes:
\begin{itemize}
\item  \textbf{TE modes:} $E_z=0$, nonzero $B_z$, with boundary condition $\partial_n B_z|_{\rm wall}=0$.
\item \textbf{TM modes:} $B_z=0$, nonzero $E_z$, with $E_z|_{\rm wall}=0$.
\end{itemize}
For a rectangular waveguide of cross-section $a\times b$, the standard solutions have
\begin{equation}
\psi_{mn}(x,y) = \cos\left(\tfrac{m\pi x}{a}\right)\cos\left(\tfrac{n\pi y}{b}\right)
\quad\text{or}\quad
\sin\left(\tfrac{m\pi x}{a}\right)\sin\left(\tfrac{n\pi y}{b}\right),
\end{equation}
leading to cutoff $\omega_{mn}^2 = \pi^2(\tfrac{m^2}{a^2} + \tfrac{n^2}{b^2})/(\mu\epsilon)$. Explicit mode functions allow computing normalization integrals (power, energy) below.
The time-averaged axial power flow in mode $\xi$ is
\begin{equation}
P(z) = \int_{A_{\rm cross}} \Re\bigl[\tfrac{1}{2}\mathbf{E}\times \mathbf{H}^*\bigr]\cdot \hat{\mathbf{z}} \,dA.
\end{equation}
Using mode functions, one shows
\begin{equation}\label{eq:power-flow}
P = \frac{1}{2\sqrt{\mu\epsilon}} \,\frac{\omega^2}{\omega_\xi^2}\,\sqrt{1 - \frac{\omega_\xi^2}{\omega^2}}\;\kappa_\xi \int_A |\psi_\xi|^2 dA,
\end{equation}
where $\kappa_\xi = \epsilon$ for TM modes or $\mu$ for TE modes. The ratio of power to stored energy $U$ yields group velocity $v_{g,\xi}$:
\begin{equation}
\frac{P}{U} = v_{g,\xi}, 
\quad
U = \frac{1}{2}\int_A (\epsilon|\mathbf{E}|^2 + \mu|\mathbf{H}|^2)\,dA.
\end{equation}

With finite wall conductivity, $k_z$ acquires a small imaginary part: 
\[
k_z \simeq k_z^{(0)} + \varepsilon_\xi + i\,\beta_\xi,
\]
where $\beta_\xi>0$ causes decay. The power decays as
\begin{equation}\label{eq:attenuation}
P(z) = P_0\,e^{-2\beta_\xi z}, 
\quad 
\beta_\xi \;=\; -\frac{1}{2P}\frac{dP}{dz}\Bigr|_{\rm ohmic}.
\end{equation}
We assume $\beta_\xi$ known (measured or from theory) and will later normalize by $e^{2\beta_\xi z}$. Experimentally, one must correct for ordinary loss to isolate oscillation.

\section{The Propagation and Oscillation of Dark Photons}
Introduce a dark U(1) gauge field $A'_\mu$ with mass $m_{A'}$, kinetically 
mixed with the visible photon $A_\mu$ via small parameter $\varepsilon$.
After field redefinitions, the relevant quadratic Lagrangian in interaction
basis $(A,A')$ can be written (in Lorenz gauge) as
\[
\mathcal{L} \;\supset\; -\tfrac14 F_{\mu\nu}F^{\mu\nu} \;-\;\tfrac14 F'_{\mu\nu}F'^{\mu\nu} \;-\;\tfrac12 m_{A'}^2 A'_\mu A'{}^\mu \;-\;\tfrac12 \varepsilon \,F_{\mu\nu}F'^{\mu\nu}.
\]
In the waveguide, the visible field $A$ obeys the massive dispersion due to
mode cutoff: effectively $m_\xi^2$ enters. The dark field $A'$ is not
confined transversely, so it propagates as a plane wave in $z$. The coupled
wave equation along $z$ for the two fields (suppressing polarization
indices, focusing on one polarization/mode) becomes:
\begin{equation}\label{eq:coupled-wave-eq}
\left[-\partial_t^2 + \partial_z^2 - 
\begin{pmatrix}
m_\xi^2 & -\varepsilon m_{A'}^2 \\
-\varepsilon m_{A'}^2 & m_{A'}^2
\end{pmatrix}
\right]
\begin{pmatrix}A(z,t)\\A'(z,t)\end{pmatrix} = 0.
\end{equation}
Here $A(z,t)=A(x,y)\,e^{i(k_z z - \omega t)}$ with transverse profile 
$A(x,y)$ normalized to the chosen mode, giving 
$-\partial_t^2 +\partial_z^2 + \partial_x^2 + \partial_y^2 \to -
\partial_t^2 + \partial_z^2 - m_\xi^2$ on $A$. For $A'$, transverse 
Laplacian is zero, giving $-\partial_t^2+\partial_z^2 - 0$; but mixing 
induces off-diagonal $-\varepsilon m_{A'}^2$ and diagonal $m_{A'}^2$ terms 
after field redefinition. Eq.~\eqref{eq:coupled-wave-eq} encapsulates how 
the effective masses and mixing produce two coupled modes that will split 
into two propagation eigenstates.

Assume harmonic time dependence $e^{-i\omega t}$ and spatial dependence
$e^{i k_z z}$. Insert 
\begin{equation}
A(z,t) = A_0\,e^{i k_z z - i\omega t}, 
\quad
A'(z,t) = A'_0\,e^{i k_z z - i\omega t}.
\end{equation}
Equation~\eqref{eq:coupled-wave-eq} becomes an algebraic condition on $k_z$ and amplitudes:
\begin{equation}
\bigl[\omega^2 - k_z^2\bigl]\begin{pmatrix}A_0\\A'_0\end{pmatrix}
=
\begin{pmatrix}
m_\xi^2 & -\varepsilon m_{A'}^2 \\
-\varepsilon m_{A'}^2 & m_{A'}^2
\end{pmatrix}
\begin{pmatrix}A_0\\A'_0\end{pmatrix}.
\end{equation}
Rearrange to
\begin{equation}
k_z^2 = \omega^2 - M^2, 
\quad\text{with}\quad
M^2 \equiv
\begin{pmatrix}
m_\xi^2 & -\varepsilon m_{A'}^2 \\
-\varepsilon m_{A'}^2 & m_{A'}^2
\end{pmatrix}.
\end{equation}
Thus the eigenvalues $m_{B_1}^2, m_{B_2}^2$ of $M^2$ determine the two propagation constants:
\begin{equation}
k_{z,B_i} = \sqrt{\omega^2 - m_{B_i}^2}, \quad i=1,2.
\end{equation}

Diagonalize $M^2$ by orthogonal rotation:
\begin{equation}\label{eq:rotation}
\begin{pmatrix}B_1\\B_2\end{pmatrix}
=
\begin{pmatrix}\cos\theta & -\sin\theta\\ \sin\theta & \cos\theta\end{pmatrix}
\begin{pmatrix}A\\A'\end{pmatrix}.
\end{equation}
The mixing angle $\theta$ satisfies
\begin{equation}
\tan 2\theta \;=\; \frac{2\varepsilon m_{A'}^2}{m_{A'}^2 - m_\xi^2}.
\end{equation}
Eigenvalues are
\begin{equation}\label{eq:eigenmasses}
m_{B_{1,2}}^2 \;=\; \frac{m_\xi^2 + m_{A'}^2 \pm \sqrt{(m_\xi^2 - m_{A'}^2)^2 + 4\varepsilon^2 m_{A'}^4}}{2}.
\end{equation}
When $m_\xi^2 \approx m_{A'}^2$ and $\varepsilon\ll1$, expand:
\begin{equation}
m_{B_{1,2}}^2 \simeq \frac{m^2 + m^2 \pm 2\varepsilon m^2}{2} = m^2 \pm \varepsilon m^2, 
\quad m\equiv m_\xi \approx m_{A'}.
\end{equation}
Thus splitting $\Delta m^2 \sim 2\varepsilon m^2$. Corresponding $k_{z,B_i} \approx \sqrt{\omega^2 - m^2 \mp \varepsilon m^2}$.  
Diagonalization identifies two propagation eigenstates $B_1, B_2$ with 
slightly different effective masses and hence different $k_z$ and
velocities. The small mixing $\varepsilon$ introduces 
the splitting needed for interference.

We prepare a pure visible photon state at the waveguide input ($z=0$): $A'(0)=0$. Using Eq.~\eqref{eq:rotation}, express initial amplitudes in eigenbasis:
\begin{equation}
A(0) = A_0, \quad A'(0)=0 
\;\implies\;
\begin{cases}
B_1(0) = \cos\theta\,A_0,\\
B_2(0) = \sin\theta\,A_0,
\end{cases}
\quad\text{so } B_2(0)/B_1(0)=\tan\theta.
\end{equation}
Each eigenstate then propagates as 
\begin{equation}
B_i(z,t) = B_i(0)\,e^{i(k_{z,B_i} z - \omega t)}.
\end{equation}
Hence the visible field at $(z,t)$ is
\begin{equation}\label{eq:A-evolution}
A(z,t) \;=\; \cos\theta\,B_1(0)\,e^{i(k_{z,B_1}z - \omega t)} \;+\; \sin\theta\,B_2(0)\,e^{i(k_{z,B_2}z - \omega t)}.
\end{equation}
The decomposition shows that the visible field is a superposition of two eigenmodes accumulating different phases $\propto k_{z,B_i} z$, leading to interference.

Compute axial Poynting flux $S_z \propto \Re[E\times H^*]\cdot\hat{z}$. Since fields carry factors from Eq.~\eqref{eq:A-evolution}, the power density contains terms $\propto |A(z,t)|^2$. After integrating over cross-section, one obtains:
\begin{equation}
P(z) = P_0\,e^{-2\beta_\xi z}\Bigl[\eta_1^2 + \eta_2^2 + 2\,\eta_1\eta_2 \cos\left((k_{z,B_1} - k_{z,B_2})z\right)\Bigr].\label{oscdeltak}
\end{equation}
Here
$\eta_1=\cos\theta$, $\eta_2=\sin\theta$ up to normalization. The factor $e^{-2\beta_\xi z}$ accounts for ordinary attenuation (assumed similar for both eigenmodes).
$\Delta k_z = k_{z,B_1} - k_{z,B_2}$.

At exact resonance ($m_\xi = m_{A'}$), $\theta = \pi/4$, so $\eta_1=\eta_2=1/\sqrt{2}$ and
\begin{equation}\label{eq:resonant-oscillation}
P(z) = P_0\,e^{-2\beta_\xi z}\Bigl[\tfrac12 + \tfrac12\cos(\Delta k_z z)\Bigr].
\end{equation}
This shows explicitly that power oscillates with spatial period $\lambda_{\rm osc}=2\pi/\Delta k_z$ on top of exponential decay. Detecting the cosine modulation is the objective.

\section{The Fourier analysis of Pulsed Signals 
and data processing for measurement}

We choose a periodic square pulse train for OTDR detection, because OTDR uses pulses to locate scattering at different $z$. Let the input temporal waveform shown in the panel c) of Fig.~\ref{fig2} be
\[
f(0,t) = A(t)\sin(\omega_0 t), 
\quad
A(t) = \mathrm{sign}\bigl(\mathrm{mod}(t,\tau) - \tfrac{\tau}{2}\bigr),
\]
with carrier frequency $\omega_0$ and pulse period $2\tau$ (modulation frequency $\Omega = \pi/\tau$, $\Omega \ll \omega_0$). 
Express $f(0,t)$ as,
\begin{equation}\label{eq:fourier-series}
f(0,t) \;\simeq\; \frac{A_0}{2}\sin(\omega_0 t)
+ \sum_{n=1}^{N_{\rm max}} C_n \sin\bigl((\omega_0 - n\Omega)t\bigr)
+ \sum_{n=1}^{\infty} C_n \sin\bigl((\omega_0 + n\Omega)t\bigr),
\end{equation}
where 
\begin{equation}
C_n = \frac{A_0}{n\pi}\sin\left(\tfrac{n\pi}{2}\right),
\quad N_{\rm max}\approx \lfloor \omega_0/\Omega\rfloor.
\end{equation}
The second sum covers lower sidebands, the third covers higher. Decomposing 
into sinusoidal components allows analyzing each frequency’s propagation
and interference separately.

\begin{figure}[htbp]
\begin{center}
\scalebox{1}{\epsfig{file=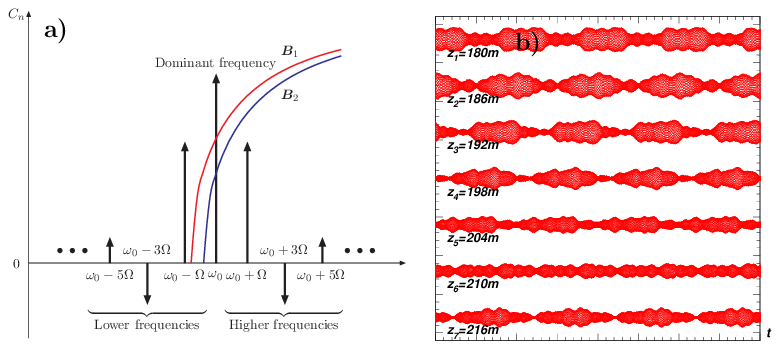}}
\caption{The sketch map of the oscillation phenomenon in the 
optical fiber. a) The component frequency of the periodic square
pulse and the resonant mode. The red and blue curve
show the group velocity of the two resonant modes;
b) The numerical simulated periodic "D+H" square pulse at different 
sample positions.}
\label{fig4}
\end{center}
\end{figure}

When $\omega_0$ is tuned close to the cutoff $\omega_\xi \approx m_{A'}$,
any component with frequency $\omega < \omega_\xi$ becomes evanescent: 
its $k_z$ is imaginary, leading to rapid exponential decay over a short 
distance. Thus, for distances beyond a small region, only components with
$\omega \ge \omega_\xi$ survive. Since $\omega_0 \approx \omega_\xi$, the
lower sidebands $(\omega_0 - n\Omega)$ fall below cutoff and die out; 
only the dominant tone $\omega_0$ and higher sidebands 
$(\omega_0 + n\Omega)$ propagate, as shown in the Fig.~\ref{fig4}.
We denote this filtered input as
\begin{equation}\label{eq:DH-tuning}
f^{\rm D+H}(0,t) = \frac{A_0}{2}\sin(\omega_0 t)
+ \sum_{n=1}^{\infty} C_n \sin\bigl((\omega_0 + n\Omega)t\bigr).
\end{equation}
Filtering simplifies analysis: we only track frequencies above cutoff, which propagate and exhibit oscillation.

For each surviving component frequency $\omega_n = \omega_0 + n\Omega$, the two eigenmodes have propagation constants:
\begin{equation}
k_{z,B_i}^{(n)} = \sqrt{\omega_n^2 - m_{B_i}^2}, \quad i=1,2.
\end{equation}
Group and phase velocities differ slightly due to $m_{B_i}$ splitting.
After traveling distance $L$, the relative phase between eigenmodes 
for component $n$ is:
\begin{equation}\label{eq:phase-difference-n}
\Delta\varphi_n = L \Bigl[\sqrt{(\omega_0 + n\Omega)^2 - m_{B_1}^2} \;-\; \sqrt{(\omega_0 + n\Omega)^2 - m_{B_2}^2}\Bigr].
\end{equation}
For small mixing ($\varepsilon \ll 1$) and 
near resonance ($\omega_0 \approx m_{B_2}\approx m_{A'}$), expand
\begin{equation}
m_{B_{1,2}}^2 \approx m^2 \pm \varepsilon m^2, 
\quad m\equiv m_{A'},
\end{equation}
so for $\omega_n \sim m$,
\begin{equation}
\sqrt{\omega_n^2 - (m^2 \pm \varepsilon m^2)} 
\approx \sqrt{\omega_n^2 - m^2} \mp \frac{\varepsilon m^2}{2\sqrt{\omega_n^2 - m^2}}.
\end{equation}
Thus their difference 
\begin{equation}
\Delta\varphi_n \approx L \,\frac{\varepsilon m^2}{\sqrt{\omega_n^2 - m^2}} \approx L \,\frac{\varepsilon m^2}{k_z^{(0)}},
\end{equation}
where $k_z^{(0)}=\sqrt{\omega_n^2 - m^2}$. At exact resonance 
$\omega_n \approx m$, $k_z^{(0)}$ small, but more careful treatment
shows leading scaling 
\begin{equation}
 \Delta\varphi_0 \sim \sqrt{2\varepsilon}\,m L\,.   \label{dltvphi0-1}
\end{equation}
as derived from full expression under 
$m_{B_{1,2}}^2 = m^2(1\pm \varepsilon)$. 

Computing $\Delta\varphi_n$ quantifies the interference period 
$\lambda_{\rm osc} = 2\pi/\Delta k_z$ and amplitude of 
oscillation for each harmonic.
At position $z$, the total field is the sum of harmonics and eigenmodes:
\begin{equation}
f^{\rm D+H}(z,t) 
= \sum_{n=0}^\infty \Bigl[\eta_1\,C'_n \sin(\omega_n t - k_{z,B_1}^{(n)} z) \;+\; \eta_2\,C'_n \sin(\omega_n t - k_{z,B_2}^{(n)} z)\Bigr],
\end{equation}
where $C'_0 = A_0/2$ for $n=0$ and $C'_n=C_n$ for $n\ge1$. The measured power (time-averaged) depends on interference term 
$\cos\left[(k_{z,B_1}^{(n)} - k_{z,B_2}^{(n)}) z\right] =
\cos\left(\Delta\varphi_n\right)$. Summing contributions effectively 
yields a modulation in the envelope of the pulse’s amplitude vs.\ $z$.
This illustrates how the waveform distortion emerges from different 
phase velocities, and more importantly, how the power oscillates.
Right panel of Fig.~\ref{fig4} shows the simulated waveform of the
D+H tune at different sample positions.

A real laser has linewidth: model each frequency component with Lorentzian distribution centered at $\omega_c$:
\begin{equation}\label{eq:lorentzian}
g(\omega,\omega_c) \;=\; \frac{1}{\pi}\frac{\Gamma^2}{(\omega - \omega_c)^2 + \Gamma^2},
\quad \Gamma = \frac{\omega_c\,\zeta}{2},\quad \zeta = \frac{\Delta\nu}{\nu}.
\end{equation}
Here $\zeta\sim10^{-5}$ (ordinary) or $\zeta\sim10^{-11}$ (ultra-stable). 
We average the phase difference for the dominant frequency component around $\omega_c \approx m_{B_2}$
\begin{eqnarray}\label{eq:avg-phase-shift}
\overline{\Delta\varphi_0} 
&=& \frac{1}{\Gamma}\int_{m_{B_2}}^{\infty} \Bigl[\sqrt{\omega^2 - (m_{B_2}^2 - 2\varepsilon m_{A'}^2)} \;-\; \sqrt{\omega^2 - m_{B_2}^2}\Bigr]\, g(\omega, m_{B_2}) \, d\omega.\nonumber\\
&=&\frac{4\varepsilon L}{\lambda}
\int_{0}^{\infty}\frac{1}{\sqrt{x\zeta+2\varepsilon}+\sqrt{x\zeta}}\frac{1}{1+x^{2}}dx\,.
\end{eqnarray}
Change variables and approximate for $\zeta \ll\varepsilon$,
We can see that $\overline{\Delta\varphi_0} $ returns to 
Eq.~~\eqref{eq:avg-phase-shift}
\begin{equation}
\overline{\Delta\varphi_0} \approx \sqrt{2\varepsilon}\,m_{A'} L,
\end{equation}
showing that finite linewidth reduces coherence only if $\zeta \gtrsim \varepsilon$. If $\zeta \gg \varepsilon$, oscillation averages out.

 Incorporating laser linewidth shows realistic sensitivity limits. It
 justifies using long-coherence lasers to probe smaller $\varepsilon$.
Set the minimal detectable phase difference 
$\Delta\varphi_0 \approx 2\pi$  over fiber length $L$. 
Using $\Delta\varphi_0 \approx \sqrt{2\varepsilon}\,m_{A'}L$, solve:
\begin{equation}
\varepsilon \approx \frac{1}{2}\Bigl(\frac{2\pi}{m_{A'}L}\Bigr)^2.
\end{equation}
Equivalently,
\begin{equation}
\log_{10}\varepsilon \approx 2\log_{10}\Bigl(\frac{2\pi}{m_{A'}L}\Bigr) - \log_{10}2.
\end{equation}
In terms of wavelength $\lambda=2\pi/m_{A'}$, $L/\lambda$ enters. For $L=10^6\,$m, $\lambda=10^{-6}\,$m, one gets $\varepsilon\sim10^{-24}$.
Provides target coupling sensitivity; shows advantage of long baseline and 
resonant enhancement ($\sqrt{\varepsilon}$ scaling).
Using averaged phase from Eq.~\eqref{eq:avg-phase-shift}, require
$\overline{\Delta\varphi_0}\sim0.01\times 2\pi$ 
to solve for $\varepsilon$ given
$\zeta$ and $L$. Plotting $\varepsilon$ vs.\ $m_{A'}$ yields exclusion 
reach (as in Fig.~\ref{fig3}). 

Generally speaking, the experimental measurements of the 
oscillation is simple. Both power flow which is measured by a
power meter, and the amplitude which is shown by the oscilloscope,
can be chosen for the display in the panel a) of Fig.~\ref{fig2}.
Suppose the power flow are measured at a series of lengths
for the prepared fiber with a radius $r_i$
\begin{eqnarray}
   {\cal M}=[ \hat{P} (z_1; r_i),~ \hat P(z_2; r_i),
   ~\cdots,~\hat P(z_n; r_i)]\,.
\end{eqnarray}
Then, every sample measurement should be normalized by the attenuation 
factor $\beta_\xi$
\begin{eqnarray}
   {\cal E}=[ \hat P(z_1; r_i)e^{2\beta_\xi z_1},
   ~\hat P(z_2; r_i)e^{2\beta_\xi z_2},~\cdots,
   ~\hat P(z_n; r_i)e^{2\beta_\xi z_n}]
   =[ P(z_1; r_i),   ~P(z_2; r_i),~\cdots,  ~P(z_n; r_i)]\,.
\end{eqnarray}
Taking the  data set ${\cal E}$  to fit the cosine function 
along the $z$ will be the evidence of the oscillation.

In addition to the monochromaticity of the laser,
there are many other different kinds of noise and disturbance, 
such the noise from the equipments, thermal environment, 
and the error of the attenuation factor, etc. These disturbances
are rather difficult for a precision measurements.
Anyway, a more precise fit for the cosine wave is 
in the data processing, especially in case of 
that the oscillation wavelength is much larger than the 
fiber length. Suppose the variation of the normalized power flow
is rather small, sightly oscillates along with the propagation distance. 
Then the phase difference can be extracted by the asymptotic 
Taylor expansion of the cosine function. For example, the power flow
are measured at the input end and the tail end, giving
the measured power flow $P(0),~P(L)$. Then, expanding the
Eq.~\eqref{oscdeltak} together with the noise $P_N$
\begin{eqnarray}
    P(0;~r_i) &=& P_0 (\eta_1^2 
    +\eta_2^2+2\eta_1\eta_2)+P_N\label{expan1}\\
    P(L;~r_i) &\simeq& P_0\left[
    \eta_1^2 +\eta_2^2+2\eta_1\eta_2\left(1-\frac{1}{2}(\Delta k_z L)^2 
    +\frac{1}{24}(\Delta k_z L)^4 \right)\right]+P_N\,. \label{expan2}
\end{eqnarray}
Here $P_N$ roughly accounts for all the noises caused by the equipments,
thermal environment, and the attenuation factor talked above. Generally
$P_N$ is kind of the resolution of the measurements which should be
independent for the propagation length $L$. And it is rather
tiny compared the input power flow $P_0$. For the slightly oscillation,
we'd better define a linear relation 
\begin{eqnarray}
    M(L;~r_i)=\frac{P(L;~r_i)}{P(0;~r_i)} \simeq 
    1 - C_1 L^2 +C_2\,,
\end{eqnarray}
in which 
\begin{eqnarray}
    C_1 = \frac{\eta_1\eta_2}{(\eta_1+\eta_2)^2 }(\Delta k_z)^2,~~~~
    C_2 = \frac{P_N}{P(0;~r_i)}\,,
\end{eqnarray}
for the numerical simulation of the data.
A non-zero $C_1$ serve as the criteria for oscillation detection, 
and the for the resonant mode with corresponding radius $r_i$ gives
us further information of the dark photon. Then the $\chi^2$ is
defined as
\begin{eqnarray}
    \chi^2=\sum_n^N\left[
    M(L_n;~r_i)- 1 + C_1 L_n^2 -C_2\right]^2\,.
\end{eqnarray}
The Least Squared Method gives the best matching result
\begin{eqnarray}
    C_1 &=& \frac{\left(\sum_n^N L_n^2\right)
    \left(\sum_n^N M(L_n;~r_i)\right)
    -N\left(\sum_n^N[M(L_n;~r_i)L_n^2]\right)}
    {N\left(\sum_n^NL_n^4\right)-\left(\sum_n^NL_n^2\right)^2}\,,\\
    C_2 &=& \frac{\left(\sum_n^NL_n^4\right)
    \left(\sum_n^N M(L_n;~r_i)\right)
    -\sum_n^N[M(L_n;~r_1)L_n^2]\left(\sum_n^N
    M(L_n;~r_i)\right)}
    {N\left(\sum_n^NL_n^4\right)-\left(\sum_n^NL_n^2\right)^2}-1\,.
\end{eqnarray}
As talked above, the optical fiber are prepared with an 
increasing radii. Only the one near the resonant mode (with a 
radius $r_r$) can give significant non-zero $C_1$. 
Then more precise measurements should be taken around the 
corresponding radius $r_r$. 

Then one may worry about that 
the mixing parameters $\eta_1,~\eta_2$ are unknown,
This can be easily steered clear of by
analyzing the power flow at two different points, say
$z=L/2,~L$. The phase difference can be easily derived form 
the expansion Eq.~\eqref{expan2}, and we can define a measurable
quantity ${\cal E}(L;~r_r)$
\begin{eqnarray}
    {\cal E}(L;~r_r)=\frac{P(L/2;~r_r)-P(0;~r_r)}{P(L;~r_r)-P(0;~r_r)}\simeq 
    \frac{1}{4}\left[1+\frac{1}{16}(\Delta k_{z}L)^{2}\right]\,.
    \label{extkz}
\end{eqnarray}
Take $\Delta k_{z}L=0.01 \times 2\pi$ for the demonstration, then
the measured $({\cal E}-0.25)\times 100 \simeq 0.006$. 
A much more precise $\Delta k_{z}L$ can be extracted by this relation if
the oscillation wavelength $\lambda_{\text{osc}}$ is too much long. 
Three measured samples are enough for the fitting for the oscillation.

\end{document}